\documentclass[runningheads]{llncs}
\usepackage[T1]{fontenc}

\usepackage[breaklinks=true]{hyperref}
\usepackage[usenames,dvipsnames]{xcolor}
\usepackage{graphicx}
\usepackage{subcaption}
\usepackage{tikz}
\usepackage{adjustbox}
\usetikzlibrary{shapes.geometric}
\usepackage{listings}
\usepackage{amsmath}
\usepackage{amssymb}
\usepackage{xspace}
\usepackage{placeins}
\usepackage{epsfig}
\usepackage{flushend} 
\usepackage{float}
\usepackage[justification=centering]{caption}
\usepackage{algorithmicx}
\usepackage[ruled]{algorithm}
\usepackage[noend]{algpseudocode}
\usepackage{ctable}
\usepackage{cite}
\usepackage{mdwlist}
\usepackage{tcolorbox}
\usepackage{tabularx}
\usepackage{array}
\usepackage{colortbl}
\usepackage{xargs}  
\usepackage[]{todonotes}
\usepackage{pgf}
\usepackage{wrapfig}
\usepackage{makecell}

\newcommandx{\bl}[2][1=]{\todo[inline,
  linecolor=LimeGreen,backgroundcolor=LimeGreen!40,bordercolor=LimeGreen,#1]{(Ben)
    #2}}
\newcommandx{\rr}[2][1=]{\todo[inline,
  linecolor=YellowOrangeGreen,backgroundcolor=YellowOrange!40,bordercolor=YellowOrange,#1]{(Riley)
    #2}}
\newcommandx{\blrr}[2][1=]{\todo[inline,
  linecolor=CadetBlue,backgroundcolor=CadetBlue!40,bordercolor=CadetBlue,#1]{(Ben
    and Riley)
    #2}}
\newcommandx{\zzrr}[2][1=]{\todo[inline,
  linecolor=Brown,backgroundcolor=Brown!40,bordercolor=Brown,#1]{(Zhen
    and Riley)
    #2}}
\newcommandx{\zzbl}[2][1=]{\todo[inline,
  linecolor=Purple,backgroundcolor=Purple!40,bordercolor=Purple,#1]{(Zhen
    and Ben)
    #2}}
\newcommandx{\srkc}[2][1=]{\todo[inline,
  linecolor=MidnightBlue,backgroundcolor=MidnightBlue!40,bordercolor=MidnightBlue,#1]{(Sanghamitra
    and Koushik)
    #2}}
\newcommandx{\pb}[2][1=]{\todo[inline,
  linecolor=MidnightBlue,backgroundcolor=MidnightBlue!40,bordercolor=MidnightBlue,#1]{(Prabal)
    #2}}
\newcommandx{\ah}[2][1=]{\todo[inline,
  linecolor=Salmon,backgroundcolor=Salmon!60,bordercolor=Salmon,#1]{(Arnd)
    #2}}
\newcommandx{\zz}[2][1=]{\todo[inline,
  linecolor=RawSienna,backgroundcolor=RawSienna!40,bordercolor=RawSienna,#1]{(Zhen)
    #2}}
\newcommandx{\all}[2][1=]{\todo[inline,
  linecolor=red,backgroundcolor=red!80,bordercolor=red,#1]{(All) #2}}

\newcommand{\modest}{\textsc{Modest}\xspace}
\newcommand{\modestToolset}{\modest \textsc{Toolset}\xspace}
\newcommand{\mcsta}{\textsc{mcsta}\xspace}
\newcommand{\modes}{\textsc{modes}\xspace}
\newcommand{\optrun}{\textit{resistiveNoise}\xspace}
\newcommand{\noiserun}{\textit{inductiveNoise}\xspace}

\usepackage{anyfontsize}

\begin{document}

%
%
\title{Probabilistic Verification for Reliability of a Two-by-Two
  Network-on-Chip System}
 \titlerunning{Probabilistic Verification for Reliability of a 2x2 NoC System}

%


 \author{Riley Roberts\inst{1}\orcidID{0000-0002-8676-3767} \and Benjamin Lewis\inst{1}\orcidID{0000-0003-2968-0233} \and  Arnd Hartmanns\inst{2}\orcidID{0000-0003-3268-8674} \and  Prabal Basu\inst{3}\orcidID{0000-0002-4860-1089} \and Sanghamitra Roy\inst{1}\orcidID{0000-0002-3927-1612} \and Koushik Chakraborty\inst{1}\orcidID{0000-0003-0228-2737} \and Zhen Zhang\inst{1}\orcidID{0000-0002-8269-9489}}

\institute{Utah State University, Logan, UT, USA\\
\email{\{riley.roberts, benjamin.lewis\}@aggiemail.usu.edu, \{koushik.chakraborty, sanghamitra.roy, zhen.zhang\}@usu.edu}\\
\and University of Twente, Enschede, The Netherlands
\email{a.hartmanns@utwente.nl} \and Cadence Design Systems, CA, USA \email{bprabal@cadence.com}
}
\authorrunning{R. Roberts et al.}


%
\maketitle              

\begin{abstract}
Modern network-on-chip (NoC) systems face reliability issues due to process and environmental variations. The power supply noise (PSN) in the power delivery network of a NoC plays a key role in determining reliability. PSN leads to voltage droop, which can cause timing errors in the NoC. This paper makes a novel contribution towards formally analyzing PSN in NoC systems. We present a probabilistic model checking approach to observe the PSN in a generic 2x2 mesh NoC with a uniform random traffic load. Key features of PSN are measured at the behavioral level. To tackle state explosion, we apply incremental abstraction techniques, including a novel probabilistic choice abstraction, based on observations of NoC behavior. The \modestToolset is used for probabilistic modeling and verification. Results are obtained for several flit injection patterns to reveal their impacts on PSN. Our analysis finds an optimal flit pattern generation with zero probability of PSN events and suggests spreading flits rather than releasing them in consecutive cycles in order to minimize PSN.




\end{abstract}
\section{Introduction}\label{sec_intro}

As the complexity advances in designing reliable distributed many-core systems, 
\emph{network-on-chip} (NoC) has become the de-facto standard for
on-chip communication. In general, their architecture composes of topologically homogeneous routers operating
synchronously in a decentralized manner, and communication is
governed by a predefined routing protocol. While sharing similarity with
conventional computer networks, a NoC design faces unique reliability
challenges, such as process and environmental variations. Precise
evaluation of the NoC early in the design flow is paramount to establish
rigorous reliability and performance guarantees. NoC reliability
analysis has to capture and quantify the design's inherent
distributive and reactive characteristics. 

Existing
literature lacks probabilistic verification of the NoC. Formal
verification of the NoC has focused primarily on 
functional correctness~\cite{SalamatKEB16, Zhang2014, Zhang2016,
  Verbeek2011, Verbeek2014}, checking performance~\cite{ZamanH14,chen2010formal,HolcombEECS2013228}, and
security~\cite{WasselGOHKCS14,SepulvedaASBS18}. Advances in
formal verification of probabilistic systems have produced mature
tools such as the \modestToolset~\cite{Hartmanns2014}, which includes the
\mcsta\ probabilistic model checker and the \modes\ statistical model
checker \cite{Budde2018}, Storm~\cite{Dehnert2017}, and
PRISM~\cite{Kwiatkowska2011}. However, these existing works
have not focused on the NoC domain.



Building on the
previous success in the probabilistic verification of a NoC central
router~\cite{Lewis2019}, we present probabilistic verification of the
\emph{power supply noise} (PSN) for a 2x2 mesh NoC system and its impact on the
 system's reliability under uniform random traffic loads. We
 cumulatively apply abstraction techniques to tackle
state space explosion for a 2x2 mesh NoC model. This includes a novel abstraction
technique based on changing probabilistic choices derived from critical
observations under design constraint assumptions. Verification results
show significant scalability of our abstraction techniques, which reduce the state space growth from exponential to
polynomial. They reveal extremely low PSN activity in
the 2x2 NoC under an every other clock cycle flit injection pattern, while
showing relatively high PSN under a burst style flit
injection. This indicates the large impact design decisions have on
PSN. Additionally, we report on an efficient flit generation pattern
that incurs \emph{zero} probability for PSN events and
make recommendations for flit generation patterns to minimize
PSN.

\section{Motivation}\label{sec:motiv}
  
PSN in the power delivery network of an integrated circuit is composed of
two major components: (a) resistive noise---the product of the current drawn
and the lumped resistance of the circuit (\emph{i$\times$r}); and (b) inductive noise---proportional to the rate of change of current through the inductance of the power grid (\emph{$\frac{\Delta i}{\Delta t}$}). For a distributed system such as a NoC, the latter plays a central role \cite{basu:iconoclast:tvlsi17}.

A high inductive noise is responsible for the intermittent peaks in the cycle-wise noise profile of a NoC. This noise is substantially growing with technology scaling. It has been recently shown that in an 8$\times$8 NoC, the peak PSN can increase from 40\% of the supply voltage at the 32-nm technology node to about 80\% of the supply voltage at the 14-nm technology node, while running a uniform-random synthetic traffic pattern \cite{basu:iconoclast:tvlsi17}. Such a droop can radically degrade the delay of various on-chip circuit components causing timing errors in the pipe-stages of the NoC routers. Hence, PSN worsens the reliability and performance of the on-chip communication.

Existing approaches to mitigate PSN are a far cry from a truly reliable NoC design paradigm that can be deployed in mission-critical systems,
as they do not guarantee the worst-case peak PSN \cite{rajesh:drnoc:islped15,basu:iconoclast:tvlsi17}. These works do not provide any
bounds on the temporal PSN profile for a router, given an application execution. Hence, temporal high peak PSNs may still exacerbate the NoC reliability across different operating conditions. To address this critical reliability challenge, we show that probabilistic verification can offer
precise bounds on the performance and reliability  of the NoC. Our proposed techniques will 
lead the way to future reliable NoC designs.

\section{Concrete Formal Model for NoC}
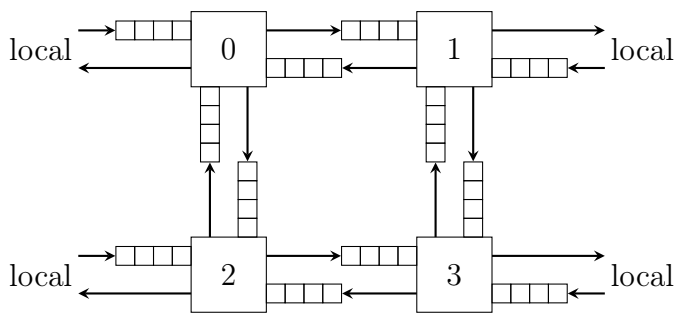
\begin{figure}
    \centering
    \begin{tikzpicture}[>=stealth, auto, scale=1, every node/.style={transform shape}]
        \node[rectangle, draw=black, minimum size=1cm] (0) at (0,3) {\large 0};
        \node[rectangle, draw=black, minimum size=1cm] (1) at (3,3) {\large 1};
        \node[rectangle, draw=black, minimum size=1cm] (2) at (0,0) {\large 2};
        \node[rectangle, draw=black, minimum size=1cm] (3) at (3,0) {\large 3};
        
        \node[] (local0) at (-2.5, 3) {\large local};
        \node[] (local1) at (5.5, 3) {\large local};
        \node[] (local2) at (-2.5, 0) {\large local};
        \node[] (local3) at (5.5, 0) {\large local};

        \node[rectangle, draw=black, minimum size=.25cm] (0ewb1) at (0.625,2.75) {};
        \node[rectangle, draw=black, minimum size=.25cm] (0ewb2) at (0.875,2.75) {};
        \node[rectangle, draw=black, minimum size=.25cm] (0ewb3) at (1.125,2.75) {};
        \node[rectangle, draw=black, minimum size=.25cm] (0ewb4) at (1.375,2.75) {};
        
        \node[rectangle, draw=black, minimum size=.25cm] (1ewb4) at (1.625,3.25) {};
        \node[rectangle, draw=black, minimum size=.25cm] (1ewb3) at (1.875,3.25) {};
        \node[rectangle, draw=black, minimum size=.25cm] (1ewb2) at (2.125,3.25) {};
        \node[rectangle, draw=black, minimum size=.25cm] (1ewb1) at (2.375,3.25) {};
        
        \node[rectangle, draw=black, minimum size=.25cm] (2ewb1) at (0.625,-.25) {};
        \node[rectangle, draw=black, minimum size=.25cm] (2ewb2) at (0.875,-.25) {};
        \node[rectangle, draw=black, minimum size=.25cm] (2ewb3) at (1.125,-.25) {};
        \node[rectangle, draw=black, minimum size=.25cm] (2ewb4) at (1.375,-.25) {};
        
        \node[rectangle, draw=black, minimum size=.25cm] (3ewb4) at (1.625,.25) {};
        \node[rectangle, draw=black, minimum size=.25cm] (3ewb3) at (1.875,.25) {};
        \node[rectangle, draw=black, minimum size=.25cm] (3ewb2) at (2.125,.25) {};
        \node[rectangle, draw=black, minimum size=.25cm] (2ewb1) at (2.375,.25) {};
        
        \node[rectangle, draw=black, minimum size=.25cm] (0nsb1) at (-0.25,2.375) {};
        \node[rectangle, draw=black, minimum size=.25cm] (0nsb2) at (-0.25,2.125) {};
        \node[rectangle, draw=black, minimum size=.25cm] (0nsb3) at (-0.25,1.875) {};
        \node[rectangle, draw=black, minimum size=.25cm] (0nsb4) at (-0.25,1.625) {};
    
        \node[rectangle, draw=black, minimum size=.25cm] (1nsb1) at (2.75,2.375) {};
        \node[rectangle, draw=black, minimum size=.25cm] (1nsb2) at (2.75,2.125) {};
        \node[rectangle, draw=black, minimum size=.25cm] (1nsb3) at (2.75,1.875) {};
        \node[rectangle, draw=black, minimum size=.25cm] (1nsb4) at (2.75,1.625) {};
        
        \node[rectangle, draw=black, minimum size=.25cm] (2nsb4) at (0.25,1.375) {};
        \node[rectangle, draw=black, minimum size=.25cm] (2nsb3) at (0.25,1.125) {};
        \node[rectangle, draw=black, minimum size=.25cm] (2nsb2) at (0.25,0.875) {};
        \node[rectangle, draw=black, minimum size=.25cm] (2nsb1) at (0.25,0.625) {};
        
        \node[rectangle, draw=black, minimum size=.25cm] (3nsb4) at (3.25,1.375) {};
        \node[rectangle, draw=black, minimum size=.25cm] (3nsb3) at (3.25,1.125) {};
        \node[rectangle, draw=black, minimum size=.25cm] (3nsb2) at (3.25,0.875) {};
        \node[rectangle, draw=black, minimum size=.25cm] (3nsb1) at (3.25,0.625) {};
        
        \node[rectangle, draw=black, minimum size=.25cm] (0lb1) at (-0.625,3.25) {};
        \node[rectangle, draw=black, minimum size=.25cm] (0lb2) at (-0.875,3.25) {};
        \node[rectangle, draw=black, minimum size=.25cm] (0lb3) at (-1.125,3.25) {};
        \node[rectangle, draw=black, minimum size=.25cm] (0lb4) at (-1.375,3.25) {};
        
        \node[rectangle, draw=black, minimum size=.25cm] (1lb1) at (3.625,2.75) {};
        \node[rectangle, draw=black, minimum size=.25cm] (1lb2) at (3.875,2.75) {};
        \node[rectangle, draw=black, minimum size=.25cm] (1lb3) at (4.125,2.75) {};
        \node[rectangle, draw=black, minimum size=.25cm] (1lb4) at (4.375,2.75) {};
        
        \node[rectangle, draw=black, minimum size=.25cm] (2lb1) at (-0.625,.25) {};
        \node[rectangle, draw=black, minimum size=.25cm] (2lb2) at (-0.875,.25) {};
        \node[rectangle, draw=black, minimum size=.25cm] (2lb3) at (-1.125,.25) {};
        \node[rectangle, draw=black, minimum size=.25cm] (2lb4) at (-1.375,.25) {};
        
        \node[rectangle, draw=black, minimum size=.25cm] (3lb1) at (3.625,-.25) {};
        \node[rectangle, draw=black, minimum size=.25cm] (3lb2) at (3.875,-.25) {};
        \node[rectangle, draw=black, minimum size=.25cm] (3lb3) at (4.125,-.25) {};
        \node[rectangle, draw=black, minimum size=.25cm] (3lb4) at (4.375,-.25) {};
        
        \draw[->, thick] (2.5,2.75) -- (0ewb4);
        \draw[->, thick] (0.5,3.25) -- (1ewb4);
        \draw[->, thick] (2.5,-.25) -- (2ewb4);
        \draw[->, thick] (0.5,.25) -- (3ewb4);
        \draw[->, thick] (-.25,0.5) -- (0nsb4);
        \draw[->, thick] (2.75,0.5) -- (1nsb4);
        \draw[->, thick] (0.25,2.5) -- (2nsb4);
        \draw[->, thick] (3.25,2.5) -- (3nsb4);
        
        \draw[->, thick] (-2, 3.25) -- (0lb4);
        \draw[->, thick] (-0.5, 2.75) -- (-2, 2.75);
        
        \draw[->, thick] (3.5, 3.25) -- (5, 3.25);
        \draw[->, thick] (5, 2.75) -- (1lb4);
        
        \draw[->, thick] (-2, 0.25) -- (2lb4);
        \draw[->, thick] (-0.5, -.25) -- (-2, -.25);
        
        \draw[->, thick] (3.5, 0.25) -- (5, 0.25);
        \draw[->, thick] (5, -.25) -- (3lb4);
        
    \end{tikzpicture}
    \caption{Architecture of the 2x2 NoC model.}
    \label{fig:2x2NoC}
  \end{figure}

This work analyzes the synchronous 2x2 mesh NoC shown in
Figure~\ref{fig:2x2NoC}. There are four symmetric routers, each with three
incoming channels: one in the horizontal X direction, one in the vertical Y direction, and a local
channel. Each channel has a buffer with the capacity of
storing four network flits. Each router has an arbiter which resolves
conflicts, i.e., multiple input flits competing for the same output
direction, and forwards the winning flit in one clock cycle. The arbiter uses
Round-Robin protocol to resolve multiple simultaneous requests to
ensure fairness in each direction. The NoC uses X-Y routing, where a
flit is first routed in the X direction until it is at the destination router or in the same column as the destination. It is then routed
in the Y direction to the destination. For example, in 
Figure~\ref{fig:2x2NoC}, if router 0 were to receive a flit destined
for router 3 on the local channel, the flit would be sent first in the
X direction to router 1, then in the Y direction to router 3.

{}
  
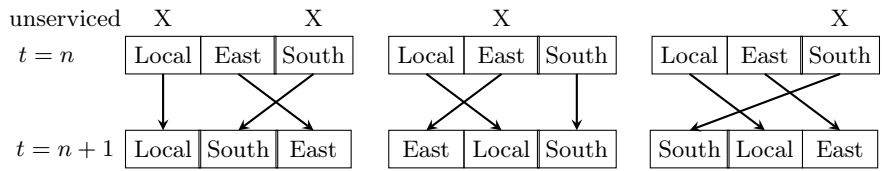
\begin{figure}
    \centering
    \begin{tikzpicture}[>=stealth, auto, scale=1, every node/.style={transform shape}]
        \node[] (unserviced) at (-1.3,.5) {unserviced};
        \node[] (n) at (-1.55,0) {$t=n$};
        \node[] (n+1) at (-1.3,-1.25) {$t=n+1$};

        \node[] (local) at (0,.5) {X};
        \node[] (south) at (2,.5) {X};
    
        \node[rectangle, draw=black, minimum height=0.5cm, minimum width=1cm] (1b1) at (0,0) {Local};
        \node[rectangle, draw=black, minimum height=0.5cm, minimum width=1cm] (1b2) at (1,0) {East};
        \node[rectangle, draw=black, minimum height=0.5cm, minimum width=1cm] (1b3) at (2,0) {South};
        
        \node[] (south) at (4.5,.5) {X};
    
        \node[rectangle, draw=black, minimum height=0.5cm, minimum width=1cm] (2b1) at (3.5,0) {Local};
        \node[rectangle, draw=black, minimum height=0.5cm, minimum width=1cm] (2b2) at (4.5,0) {East};
        \node[rectangle, draw=black, minimum height=0.5cm, minimum width=1cm] (2b3) at (5.5,0) {South};
        
        \node[] (east) at (9,.5) {X};
    
        \node[rectangle, draw=black, minimum height=0.5cm, minimum width=1cm] (3b1) at (7,0) {Local};
        \node[rectangle, draw=black, minimum height=0.5cm, minimum width=1cm] (3b2) at (8,0) {East};
        \node[rectangle, draw=black, minimum height=0.5cm, minimum width=1cm] (3b3) at (9,0) {South};

        \node[rectangle, draw=black, minimum height=0.5cm, minimum width=1cm] (1b1') at (0,-1.25) {Local};
        \node[rectangle, draw=black, minimum height=0.5cm, minimum width=1cm] (1b2') at (1,-1.25) {South};
        \node[rectangle, draw=black, minimum height=0.5cm, minimum width=1cm] (1b3') at (2,-1.25) {East};
    
        \node[rectangle, draw=black, minimum height=0.5cm, minimum width=1cm] (2b1') at (3.5,-1.25) {East};
        \node[rectangle, draw=black, minimum height=0.5cm, minimum width=1cm] (2b2') at (4.5,-1.25) {Local};
        \node[rectangle, draw=black, minimum height=0.5cm, minimum width=1cm] (2b3') at (5.5,-1.25) {South};
    
        \node[rectangle, draw=black, minimum height=0.5cm, minimum width=1cm] (3b1') at (7,-1.25) {South};
        \node[rectangle, draw=black, minimum height=0.5cm, minimum width=1cm] (3b2') at (8,-1.25) {Local};
        \node[rectangle, draw=black, minimum height=0.5cm, minimum width=1cm] (3b3') at (9,-1.25) {East};
        
        \draw[->, thick] (0, -0.25) -- (0, -1);        
        \draw[->, thick] (1, -0.25) -- (2, -1);
        \draw[->, thick] (2, -0.25) -- (1, -1);

        \draw[->, thick] (3.5, -0.25) -- (4.5, -1);        
        \draw[->, thick] (4.5, -0.25) -- (3.5, -1);
        \draw[->, thick] (5.5, -0.25) -- (5.5, -1);
        
        \draw[->, thick] (7, -0.25) -- (8, -1);        
        \draw[->, thick] (8, -0.25) -- (9, -1);
        \draw[->, thick] (9, -0.25) -- (7, -1);               
    \end{tikzpicture}
    \caption{Three priority queue examples.}
    \label{fig:Priority_Queue}
  \end{figure}

When a new flit is generated in each router's local channel, its destination is uniformly randomly selected
among the other three routers. The routers then route all flits
simultaneously. First, a flit's next forwarding direction is
determined by the X-Y routing protocol by comparing its
destination and the current router. 
The arbiter then forwards the flit to the neighboring router and
resolves conflicts if they arise. We use a priority queue in the arbiter to implement the Round-Robin scheduling
mechanism in order to maintain fairness when resolving
conflicts. Figure~\ref{fig:Priority_Queue} illustrates how the priority queue works. When two 
channels conflict, the one closer to the front of the priority
queue is serviced by the arbiter and the other is marked as unserviced. A channel that fails to send due to the receiving buffer being
full is also marked as unserviced. At the end of the current clock cycle,
the priority queue updates by shifting all unserviced channels to the
front of the queue and pushing those serviced to the end, while
maintaining their relative ordering. We use the high-level formal modeling language \modest{}~\cite{Hartmanns2014} to specify the
probabilistic NoC model in Figure~\ref{fig:2x2NoC}. 

Data types of the concrete NoC model are shown in
Listing~\ref{lst:datatypes}. The buffers are modeled as a FIFO queue
with a capacity of four. Each channel is modeled as a datatype containing a buffer, the
channel's priority, the forwarding direction for the flit at the front
of the queue, the ID of the channel, and a Boolean variable indicating whether the 
channel was serviced or not. Each router is modeled as a datatype with
an array of three channels, the order of which determines the
priority, and two counters \lstinline{unserviced} and \lstinline{totalUnsercived} that
keep track of the number of unserviced channels and are used by the
arbiter. The NoC model has an array of four routers.

\begin{lstlisting}[caption= Channel and router datatypes., label={lst:datatypes}]
datatype channel = {int direction, 
         int id, bool serviced, int priority, 
         queue buffer};		
datatype router = {int unserviced, 
         int totalUnserviced, channel[] channelArray};
\end{lstlisting}

The two components of PSN are modeled as follows. Resistive noise is
measured by accumulating the clock cycles where there is 
high router activity, i.e., all three buffers in a router are able
to forward flits. This is represented by a variable \optrun, which
increments every time a router encounters a cycle with high
activity. Inductive noise is measured by accumulating clock cycles where there is a
high rate of change of current draw. This directly corresponds to
cycles where a router switches between forwarding all flits to forwarding no flits
and vice versa. Represented by the variable \noiserun, it
increments on the cycles of abrupt change in router activity. 

The relation between the two PSN components, \optrun and \noiserun, and
their associated real-world applications 
is discussed in~\cite{basu:iconoclast:tvlsi17,dahir:NoC_PSN:TC14}. We rely on the analysis presented there as the foundation. The
purpose of specifying and checking these properties is to understand
the likelihood of PSN at a behavioral level under a given routing
protocol. The intermittent peaks in the cycle-wise noise profile of
the NoC is strongly correlated to the NoC router activities. Hence,
understanding PSN behaviorally can help with the design of routing
protocols and other higher-level NoC designs independent of the
physical hardware implementation of the NoC.


\section{Need for Abstraction}~\label{sec:abstraction}

The state spaces of all probabilistic formal models presented in this work are large \emph{discrete-time Markov chains} (DTMC), and they are analyzed by the tools in the \modestToolset, namely, \mcsta\ and \modes. The concrete model presented in the previous section incurs an exponential state space growth as the number of considered clock cycles increases. This is due to the combinations of flits in all twelve buffers in the model. To address this issue, we investigate several abstraction techniques. These abstraction techniques are performed \emph{cumulatively}: results labeled using the name of one technique are the results of that technique applied on top of all previous abstractions.


\subsection{Predicate Abstraction to Simplify Complex Data Structures} 


Predicate abstraction is applied first to the concrete model. It works by formulating predicates that
capture critical decision points in the concrete model, and then
transforming the model to include only predicate variables. In our model, the predicate abstraction converts all complex data structures into predicate variables. For example, the two predicate variables in Listing~\ref{lst:predicates} are defined as the two possible forwarding directions of the front flit in the local channel buffer of router 0. Performing predicate abstraction in this manner significantly simplifies the complex data structures in the concrete model, and it also preserves the properties \optrun and \noiserun, since router and arbiter activity remain unchanged after abstraction. Note that predicate abstraction is only applied to simplify complex data structures in our concrete model to turn them into predicate variables, instead of the entire concrete model. Therefore, it does not introduce nondeterministic behavior as a result. Unfortunately, the predicate abstracted model still incurs significant state space explosion with 7 or more clock cycles. 
\begin{lstlisting}[caption=Example predicate variables after predicate abstraction., label={lst:predicates}]
bool r0L1;//noc[0].channel[local].direction==east
bool r0L2;//noc[0].channel[local].direction==south
\end{lstlisting}


\subsection{Probabilistic Choice Abstraction}

Next, we present a novel \emph{probabilistic choice abstraction} technique. The idea comes from the following observation. The flit's destination is selected by a uniform random distribution when it is input into the local buffer of each router, but the destination information is not checked \emph{until} the flit enters the router, where it is used to decide the flit forwarding direction. This implies that, when generating states, enumeration of all possible values for the destination variable does not need to happen at its initial assignment, but can be delayed until the location where its value is first checked. Furthermore, the destination variable is not in use beyond this point. Consequently, this variable can be entirely replaced by a probabilistic choice when its value is evaluated, i.e., when the router decides the flit's direction, while preserving the model behavior. The noteworthy state reduction comes from delaying the enumeration of all of the possible values until it is evaluated. This idea is similar to the concept of \emph{constant propagation} in compiler optimization.


\begin{figure}[tbhp]
    \scalebox{0.65}{
    \begin{tikzpicture}
    \node[rectangle, draw=black, text width=1.25cm] (0) at(-6,3.5) {$z=1\newline x=0\newline y=0\newline \mathit{clk}=0$};
    \node[rectangle, draw=black, text width=1.25cm] (21) at(-11,-0.5) {$z=1\newline x=1\newline y=1\newline \mathit{clk}=1$};
    \node[rectangle, draw=black, text width=1.25cm] (22) at(-9,-0.5) {$z=1\newline x=1\newline y=2\newline \mathit{clk}=1$};
    \node[rectangle, draw=black, text width=1.25cm] (23) at(-7,-0.5) {$z=1\newline x=2\newline y=1\newline \mathit{clk}=1$};
    \node[rectangle, draw=black, text width=1.25cm] (24) at(-5,-0.5) {$z=1\newline x=2\newline y=2\newline \mathit{clk}=1$};
    \node[rectangle, draw=black, text width=1.25cm] (25) at(-3,-0.5) {$z=1\newline x=3\newline y=1\newline \mathit{clk}=1$};
    \node[rectangle, draw=black, text width=1.25cm] (26) at(-1,-0.5) {$z=1\newline x=3\newline y=2\newline \mathit{clk}=1$};
    
    \node[rectangle, draw=black, text width=1.25cm] (31) at(-11,-4.5) {$z=1\newline x=1\newline y=1\newline \mathit{clk}=2$};
    \node[rectangle, draw=black, text width=1.25cm] (32) at(-9,-4.5) {$z=0\newline x=1\newline y=2\newline \mathit{clk}=2$};
    \node[rectangle, draw=black, text width=1.25cm] (33) at(-7,-4.5) {$z=2\newline x=2\newline y=1\newline \mathit{clk}=2$};
    \node[rectangle, draw=black, text width=1.25cm] (34) at(-5,-4.5) {$z=1\newline x=2\newline y=2\newline \mathit{clk}=2$};
    \node[rectangle, draw=black, text width=1.25cm] (35) at(-3,-4.5) {$z=2\newline x=3\newline y=1\newline \mathit{clk}=2$};
    \node[rectangle, draw=black, text width=1.25cm] (36) at(-1,-4.5) {$z=2\newline x=3\newline y=2\newline \mathit{clk}=2$};
    
    \draw[->] (-6,5) -- (0);
    
    \draw[->] (0) to [out=180, in=90] node[left,pos=0.9]{\Large{\textcolor{blue}{$\mathbf{\frac{1}{6}}$}}} node[left, yshift =0.5cm, text width=1.1cm]{$\mathit{clk}{++},\newline x \! := \! 1,\newline y \! := \! 1$}(21);
    \draw[->] (0) to[out=200, in=75] node[left,pos=0.9]{\Large{\textcolor{blue}{$\mathbf{\frac{1}{6}}$}}} node[left, xshift=-0.2cm, text width=1.1cm]{$ \mathit{clk}{++}, \newline x \! := \! 1, \newline y \! := \! 2$}(22);
    \draw[->] (0) -- node[right,pos=0.85]{\Large{\textcolor{blue}{$\mathbf{\frac{1}{6}}$}}} node[left, xshift = -.2cm, text width=1cm]{$\mathit{clk}{++}, \newline x \!\! := \!\! 2,\newline y \!\! := \!\! 1$}(23);
    \draw[->] (0) -- node[left,pos=0.85]{\Large{\textcolor{blue}{$\mathbf{\frac{1}{6}}$}}} node[right, xshift=0.1cm, text width=1cm]{$\mathit{clk}{++}, \newline x \!\! := \!\! 2,\newline y \! :=\!\! 2$}(24);
    \draw[->] (0) to[out=340, in=105] node[right,pos=0.9]{\Large{\textcolor{blue}{$\mathbf{\frac{1}{6}}$}}} node[right, xshift=0.3cm, yshift = -0.05cm, text width=1.1cm]{$\mathit{clk}{++}, \newline x \! := \! 3, \newline y \! := \! 1$}(25);
    \draw[->] (0) to [out=0, in=90] node[right,pos=0.9]{\Large{\textcolor{blue}{$\mathbf{\frac{1}{6}}$}}} node[right, yshift = 0.5cm, xshift = 0.4cm, text width=1.1cm]{$\mathit{clk}{++}, \newline x \!:= \!3, \newline y \! := \! 2$}(26);
    
    \draw[->] (21) -- node[left, text width = 1.25cm]{$\mathit{clk}{++},\newline x\!\! = \!\! y \rightarrow z \! := \! z$} (31);
    
    \draw[->] (22) -- node[left, text width = 1.25cm]{$\mathit{clk}{++}, \newline x \!\! < \!\! y \rightarrow z \!\! := \!\! z-1$} (32);
    
    \draw[->] (23) -- node[left, text width = 1.25cm]{$\mathit{clk}{++}, \newline x \! \! > \! \! y \rightarrow z \!\! := \!\! z+1$} (33);
    
    \draw[->] (24) -- node[left, text width = 1.25cm]{$\mathit{clk}{++}, \newline x \!\! = \! \!  y \rightarrow z \! := \! z$} (34);
    
    \draw[->] (25) -- node[left, text width = 1.25cm]{$\mathit{clk}{++}, \newline x \!\! > \! \! y \rightarrow z \!\! := \!\! z+1$} (35);
    
    \draw[->] (26) -- node[left, text width = 1.25cm]{$\mathit{clk}{++}, \newline x \!\! > \! \! y \rightarrow z \!\! := \!\! z+1$} (36);
  \end{tikzpicture}}
  \scalebox{0.65}{
    \begin{tikzpicture}
    \node[rectangle, draw=black, text width=1.25cm] (start) at(6,2.5)             {$z=1 \newline \mathit{clk}=0$};
    
    \node[rectangle, draw=black, text width=1.25cm] (0) at(6,-0.5)             {$z=1 \newline \mathit{clk}=1$};
    
    \node[rectangle, draw=black, text width=1.25cm] (1) at(3.5,-3.75) {$z=0 \newline \mathit{clk}=2$};
    \node[rectangle, draw=black, text width=1.25cm] (2) at(6,-3.75) {$z=1 \newline \mathit{clk}=2$};
    \node[rectangle, draw=black, text width=1.25cm] (3) at(8,-3.75) {$z=2 \newline \mathit{clk}=2$};
    
    \draw[->] (6,4) -- (start);
    \draw[->] (start) --node[left]{$\mathit{clk}++$} (0);
    \draw[->] (0) -- node[above,pos=0.4]{\Large{\textcolor{blue}{$\mathbf{\frac{1}{6}}$}}} node[left, text width = 1.5cm, xshift = -0.1cm, yshift=.25cm]{$\mathit{clk}{++}, \newline z \! := \! z-1$} (1);
    \draw[->] (0) -- node[right,pos=0.25]{\Large{\textcolor{blue}{$\mathbf{\frac{1}{3}}$}}} node[left, text width = 1.1cm, xshift = 0.15cm, yshift=-0.25cm]{$\mathit{clk}{++}, \newline z \! := \! z$} (2);
    \draw[->] (0) -- node[above,pos=0.4]{\Large{\textcolor{blue}{$\mathbf{\frac{1}{2}}$}}} node[right, text width = 1.5cm, xshift = 0.2cm, yshift=0.25cm]{$\mathit{clk}{++}, \newline z \! := \! z+1$} (3);
  \end{tikzpicture}}
\caption{Probabilistic Choice Abstraction}
\label{fig:probabilistic_abstraction_example}
\end{figure}
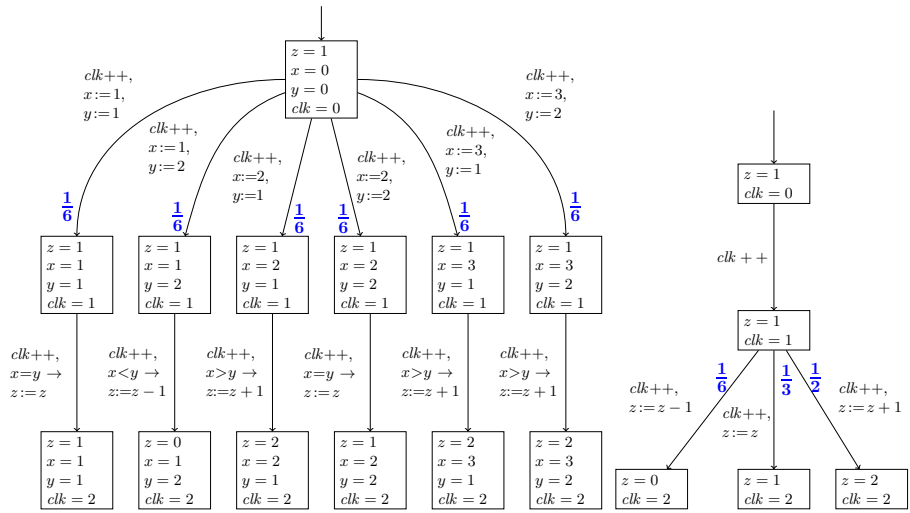

Figure~\ref{fig:probabilistic_abstraction_example} compares state graphs for an illustrative model before and after applying probabilistic choice abstraction. The left one is the fully expanded state space of a concrete model, which determines the update of $z$ by comparing two uniform random integer variables $x \in [1, 3]$ and $y \in [1, 2]$. This results in six unique states when the clock reaches 2. Variables $x$ and $y$ are only compared at this row, but not evaluated in any of the six states in the middle row, and are not used after the comparison. Thus, we can apply probabilistic choice abstraction by replacing $x$ and $y$ with a probabilistic choice with explicit probabilities as the state graph on the right illustrates. The abstract model creates only five states and preserves the behavior of the observable variables $z$ and $clk$.
 
Figure~\ref{fig:probabilistic_abstraction_example} is similar to our predicate-abstracted NoC model. In this figure, the observable variables are $z$ and $clk$. Variables $x$ and $y$ are just used to determine the behavior of $z$. Similarly in our model, the observable variables are \optrun, \noiserun and \textit{clock}. All other variables are used to determine the update behavior of \optrun and \noiserun in between clock cycles. The destinations of the flits are the only probabilistically updated variables in the model. Using this abstraction, we are able to remove the destination variables from the state space, and replace them with a probabilistic choice. This abstraction preserves the behavioral activities of the flits, which determine \optrun and \noiserun. We are currently working on developing theories and proofs for this abstraction technique. We have experimentally observed that for as many cycles as the concrete and abstract models can run, the behaviors of \optrun and \noiserun are identical across the concrete, predicate abstract, and probabilistic choice abstract models.

Probabilistic choice abstraction can be directly applied to a formal model without first generating the state graph of the concrete model. This idea is applied to the NoC model in the following way. Any flit coming from the local channel in router 0 can be destined for router 1, 2, or 3 as shown in Figure~\ref{fig:2x2NoC}. Given that the NoC uses X-Y routing and uniform random distribution of destinations, if the destination is router 1 or 3, the flit is forwarded east with probability $\frac{2}{3}$. If the destination is router 2, the flit is forwarded south with probability $\frac{1}{3}$. Any incoming flit coming from the east buffer of router 0 must have been generated in router 1 and is destined for either router 0 or router 2 with equal probability $\frac{1}{2}$. Similarly, any incoming flit in the south buffer of router 0 is generated by either router 2 or 3, and can only be destined for this router. Due to the symmetric nature of the NoC, this same pattern is true for all routers. Because the exact destination of the flit is no longer required knowledge, the buffers can be further abstracted as detailed in the next section.

\subsection{Boolean Queue Abstraction}\label{sec:BooleanQueueAbs}
With three buffers in each router, there are a total of six possible orders that determine the priority in servicing each buffer in the case of conflict. Our analysis indicates that, for the 2x2 NoC with X-Y routing, the north/south buffer can \emph{never} be in conflict with the local buffer. Therefore, the model only needs to keep track of the local and north/south buffers priority relative to the east/west buffer but not to each other. This allows us to abstract the six possible orders into four,  further reducing the state space. 

Further, because all buffers operate as FIFOs, an empty buffer element can only exist either after a non-empty element if the buffer is neither full nor empty, or at the front of the buffer if the buffer is empty. Rather than keep track of the buffer's contents, we only need the length of flit occupancy in a buffer, i.e., the number of non-empty elements. This length is represented as a bounded integer ranging between 0 and 4. In general, a buffer storing maximally $n$ Boolean variables incurs $2^n$ states, but only $n+1$ states if its occupancy is recorded. In order to maintain correct behavior through this abstraction, the four arbiter processes were synchronized on one action, rather than run sequentially.

Figure~\ref{fig:absModelDecisions} depicts the decision procedure in a router after Boolean queue abstraction. It first checks for a possible conflict by checking if the east/west buffer is empty. If the buffer is not empty (i.e., $\text{\lstinline{ewLen}} \neq 0$), the decision procedure goes into one of four branches, depending on whether the two Boolean variables \lstinline{localPriority} for the local buffer and \lstinline{nsPriority} for the north/south buffer have priority over the east/west buffer. The router model then makes a decision based on the Boolean lock variables. If the local buffer tried to send in the north/south direction, but couldn't due to a conflict or a full buffer, the lock variable \lstinline{localLns} would become \lstinline{true}, ensuring that the flit is sent in the same direction next cycle. The probabilistic decisions have probability values marked on the edges. The router then tries to service the buffers in the order listed in the figure. If a buffer fails to win the conflict resolution in sending a flit, it is locked in the same direction and its priority is advanced. Accumulation of \optrun is labeled by \lstinline{RNoise++} when all three buffers send their flits in a cycle. Figure~\ref{fig:absModelDecisions} depicts branches 1 and 2 of the four branches. Branch 3 is identical to branch 1, except the order of buffer updates changes from ``local, ns, ew'' to ``ns, ew, local''. Similarly, branch 4 changes branch 2's order from ``ew, local, ns'' to ``local, ew, ns''.
\begin{figure}[tbhp]
\centering
\input{behavior.tikz}
\caption{Decision procedure for the router model after Boolean queue abstraction.}
\label{fig:absModelDecisions}
\end{figure}


\section{Results}

All formal models described in this work have been formulated at a higher behavioral level. Therefore, it is necessary to map circuit-level behavior characterizing PSN onto the same behavioral level as these models, in order to quantitatively check PSN related properties.  As a major source of PSN, the inductive noise is proportional to
the rate of electrical current change in the circuit: abrupt router activity changes in consecutive cycles directly lead to a high rate of current change~\cite{basu:iconoclast:tvlsi17}. A high router activity is characterized as an arbiter servicing all three buffers in a cycle, while a low router activity is when the arbiter services no buffers. PSN is therefore reflected accurately by the frequency of
both high-to-low and low-to-high activities over a given timespan. All formal models described in this paper are available on Github\footnote{\url{https://github.com/formal-verification-research/Modest-Probabilistic-Models-for-NoC}}. Assuming uniform random flit destination generation at each router, we report verification results for flit injection at the rate of one flit every two clock cycles first, followed by a bursty mode injection. We then report findings of an optimal flit injection pattern that minimizes PSN. All results presented in this section were generated on a machine with an AMD Ryzen Threadripper 12-Core 3.5 GHz Processor and 132 GB memory, running Ubuntu Linux v18.04.3.

We consider the following two transient probabilistic
properties: (1)~the probability that the number of \optrun is lower-bounded by a constant $K \in \mathbb{Z}^+$ within $n$ cycles; and (2)~the probability that the total number of \noiserun is lower-bounded by a constant $K \in \mathbb{Z}^+$ within $n$ cycles. High router activity, characterized by property (1), is a key indicator of local congestion in the network, and a highly congested network leads to high PSN due to an unbalanced power density ~\cite{dahir:NoC_PSN:TC14}. Property (2) reflects an abrupt load change in a router that causes
a large inductive drop in the power delivery network~\cite{basu:iconoclast:tvlsi17}.

\subsection{Every Other Cycle Flit Injection}

\subsubsection{State Reduction from Abstraction}
\begin{figure}
    \centering
     \includegraphics[width=\textwidth]{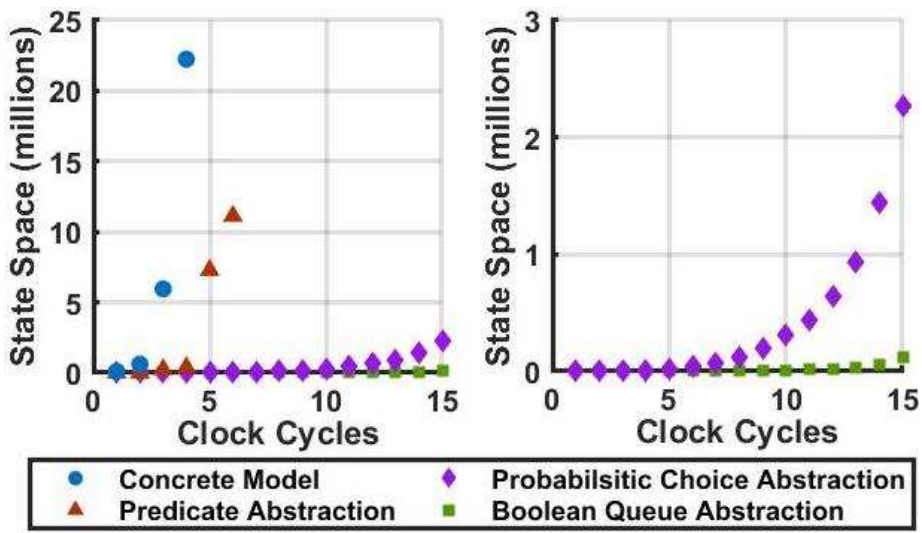}
    \caption{State count comparison.}
    \label{fig:stateCntComp}
  \end{figure}

Figure~\ref{fig:stateCntComp} illustrates the impact of applying the aforementioned abstraction techniques on state growth. The exponential cycle-wise growth of the concrete model dramatically reduces to a polynomial growth after Boolean queue abstraction, with other abstraction methods in between. Note that state growth for the presented abstraction techniques are cumulative: probabilistic choice abstraction is applied after predicate abstraction, and it is the base for Boolean queue abstraction. The exponential state growth of the concrete model is due to the probabilistic input. Every other clock cycle, four new flits are generated each with three different possible destinations. This means that if variable $x$ were to increase every other clock cycle, the states, due to just this combination, would be $3^{4x}$. This analysis is confirmed by the $R^2$ values of regression of the state space growth. The $R^2$ value for the exponential regression is close to 1 for the concrete and predicate abstract models, and the $R^2$ value for polynomial regression is closest to 1 for both probabilistic choice abstraction and Boolean queue abstraction.

In addition, we also performed exponential and polynomial regression on the state growth of the probabilistic choice and Boolean queue abstractions using only the first ten values to see how accurately each regression would predict the next five values. The polynomial regression was a more accurate prediction of the next five values than the exponential regression.
Figure~\ref{fig:stateCntComp} also illustrates the effectiveness of the cumulative abstraction steps detailed in Section~\ref{sec:abstraction}. No property verification results could be obtained on the concrete model regarding the \optrun property, as it is impossible for at least one \optrun to occur within the 4 to 5 clock cycles before state explosion occurs. The same reasoning applies to the \noiserun property. 

We keep an integer variable $clk$ as a cycle counter, which contributes to the state growth. On the other hand, by gradually increasing its upper bound, state space generation is manageable. This is because state generation only needs to represent model behaviors up to the upper bound of $clk$. 

\subsubsection{Probabilistic Model Checking (PMC)}
\begin{figure}
    \centering
     \includegraphics[width=\textwidth]{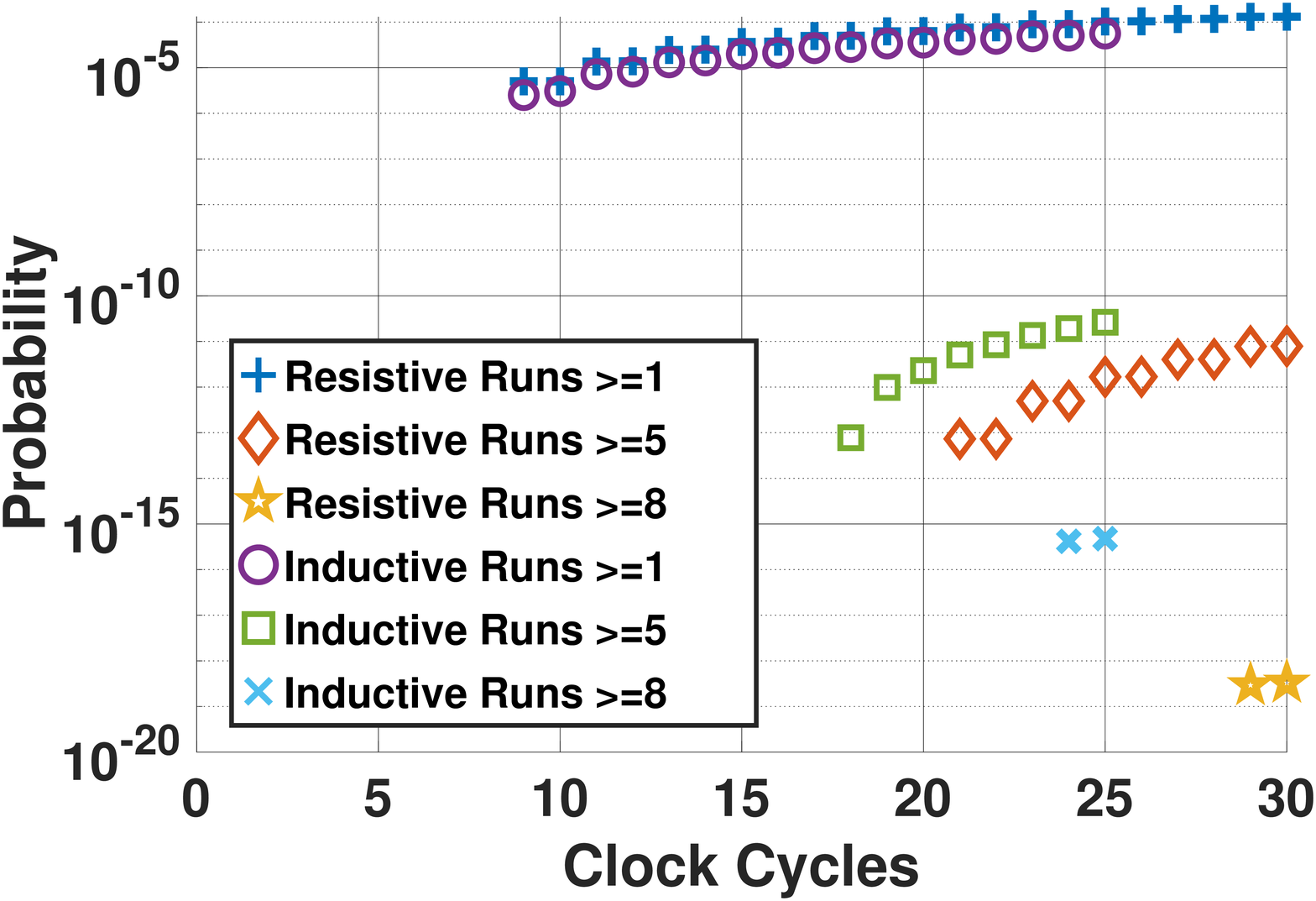}
    \caption{Probabilities for \optrun and \noiserun in every other cycle input configuration}
    \label{fig:OptResLog}
\end{figure}

We used \mcsta , an explicit-state probabilistic model checker, to calculate the probabilities of \optrun and \noiserun occurring within a given bounded number of clock cycles using the final Boolean queue abstraction model.
The results are shown in 
Figure~\ref{fig:OptResLog}.
The more abstract models perform an increasing amount of ``work''---calculating, communicating, and updating transient and state variables---on every transition.
Thus as state space explosion is alleviated, the runtime for state space exploration~rises. An attempt to run the Boolean queue abstraction model for 35 clock cycles ran for 22 hours and generated 150.5 million states. When \mcsta attempted to merge these states, it failed due to a segmentation fault indicating an out-of-memory error.

Due to the inability of the concrete model to produce any verification results, it is impossible to compare them with abstract models. Comparing the \optrun property checking results between the probabilistic choice delay and Boolean queue abstract models, a difference of \texttt{1E-7}, i.e., a $0.15\%$ difference, starts to manifest after 20 clock cycles. It is possible that this difference is due to floating-point error in the transition probabilities, which can be complicated due to the probabilistic behavior of all four arbiters being compounded. These small rounding errors can then accumulate over several clock cycles. With fewer state transitions in the Boolean queue abstracted model, it is possible that the floating-point errors accrue at a different pace. Further research is planned to investigate the minor difference.

Binary decision diagrams (BDDs) have been highly successful in hardware verification~\cite{CG18}.
We thus also explored the use of BDD-based symbolic model checking for our NoC models by exporting them to the \textsc{Jani} model interchange format~\cite{BDHHJT17} and applying the Storm model checker's \textit{hybrid} and \textit{dd} engines.
Unfortunately, with both supported BDD libraries---\textsc{Cudd} and \textsc{Sylvan}~\cite{DP17}---Storm quickly ran out of memory.
This may be due to probabilistic models requiring multi-terminal BDDs to store probability values, which often does not lead to a memory-efficient representation.

\subsubsection{Statistical Model Checking (SMC)}
Since our models represent DTMC, we can also apply the Monte Carlo simulation-based statistical model checking technique.
In contrast to PMC, it avoids state space explosion entirely, but only delivers statistical guarantees and is problematic for rare events.
We checked the \optrun property at a higher number of cycles using SMC with \modes. Running 10000 simulations, \modes reported $643$ had at least one optimal run, while $345$ had at least one noise run, after 10000 clock cycles. These values correspond to statistical probabilities of $0.0643$ and $0.0345$ respectively. Due to the rarity of these events, statistical analysis may not provide accurate results.

\subsection{Burst Flit Injection}

To more accurately model packet injection patterns in real-world applications and measure their impact on PSN, we gathered verification results for a bursty mode packet injection, where for every 10 cycles, the first 3 consecutive cycles each have a flit injected, followed by 7 idle cycles in all four routers. A surprising observation is that this input pattern drastically increases the scalability to allow verification for significantly longer clock cycles. Our analysis indicates that the NoC empties out flits completely within the 7 idle cycles before it receives the next burst of 3 flits. Other than the priority orderings, the NoC has completely reset itself to the initial condition by the end of the idle cycles. Consequently, the number of reachable states is significantly reduced, compared to the every other cycle injection pattern. This reduction leads to the generation of the entire state space for the 2x2 NoC model whereas the clock variable $clk$ is no longer used to forcibly stop the model checking after a certain number of cycles as in the every other cycle packet injection pattern. A simple 0 to 9 counter was added to maintain the 10-cycle injection pattern. Consequently, we made $clk$ a \emph{transient} variable in order to continue to check properties in relation to clock cycles. A transient variable  is only ``live'' during the assignments execution when taking a transition, which excludes it from the state vector. In this way, clock cycle progress becomes a \emph{reward} annotation to certain transitions instead of being encoded in the structure of an expanded state space. We can then formalize properties (1) and (2) as reward-bounded reachability queries:\\[6pt]
\centerline{$
\begin{aligned}
\text{(1)}\quad &\mathbf{P}_{=?}(\diamond^{[\mathrm{accumulate}(\mathit{clk}) \leqslant N]}\: \texttt{\optrun} \geqslant K)\\
\text{(2)}\quad &\mathbf{P}_{=?}(\diamond^{[\mathrm{accumulate}(\mathit{clk}) \leqslant N]} \: \texttt{\noiserun} \geqslant K)
\end{aligned}
$}\\[6pt]

Probabilistic model checking of the bursty mode with \mcsta scales to hundreds of clock cycles. In order to prevent 
the infinite accumulation of \optrun, it is upper bounded by 20, resulting in a total of 16,581,401 reachable states explored in about 3.5 hours. 
Similarly, \noiserun is upper-bounded by 8, resulting in 10,251,017 states generated in 2.16 hours. 
A smaller upper bound is required due to the addition of helper variables for accurately tracking the high-to-low and low-to-high activities for \noiserun.

\begin{figure}
\begin{subfigure}[b]{6cm}
    \centering
    \scalebox{1.1}{
     \includegraphics[width=\textwidth]{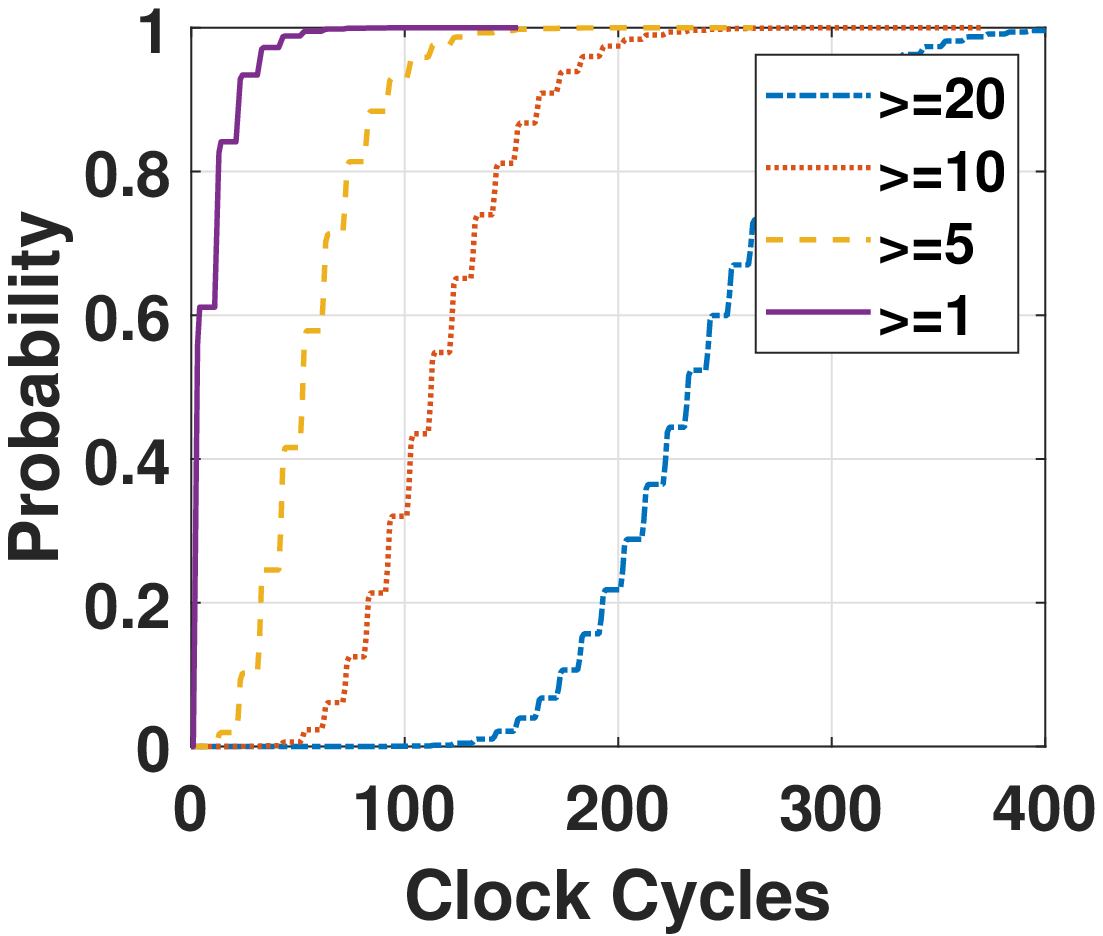}
    }
    \caption{CDF for \optrun}
\end{subfigure}
\hspace{0.1cm}
%
\begin{subfigure}[b]{6cm}
    \centering
    \scalebox{0.95}{
     \includegraphics[width=\textwidth]{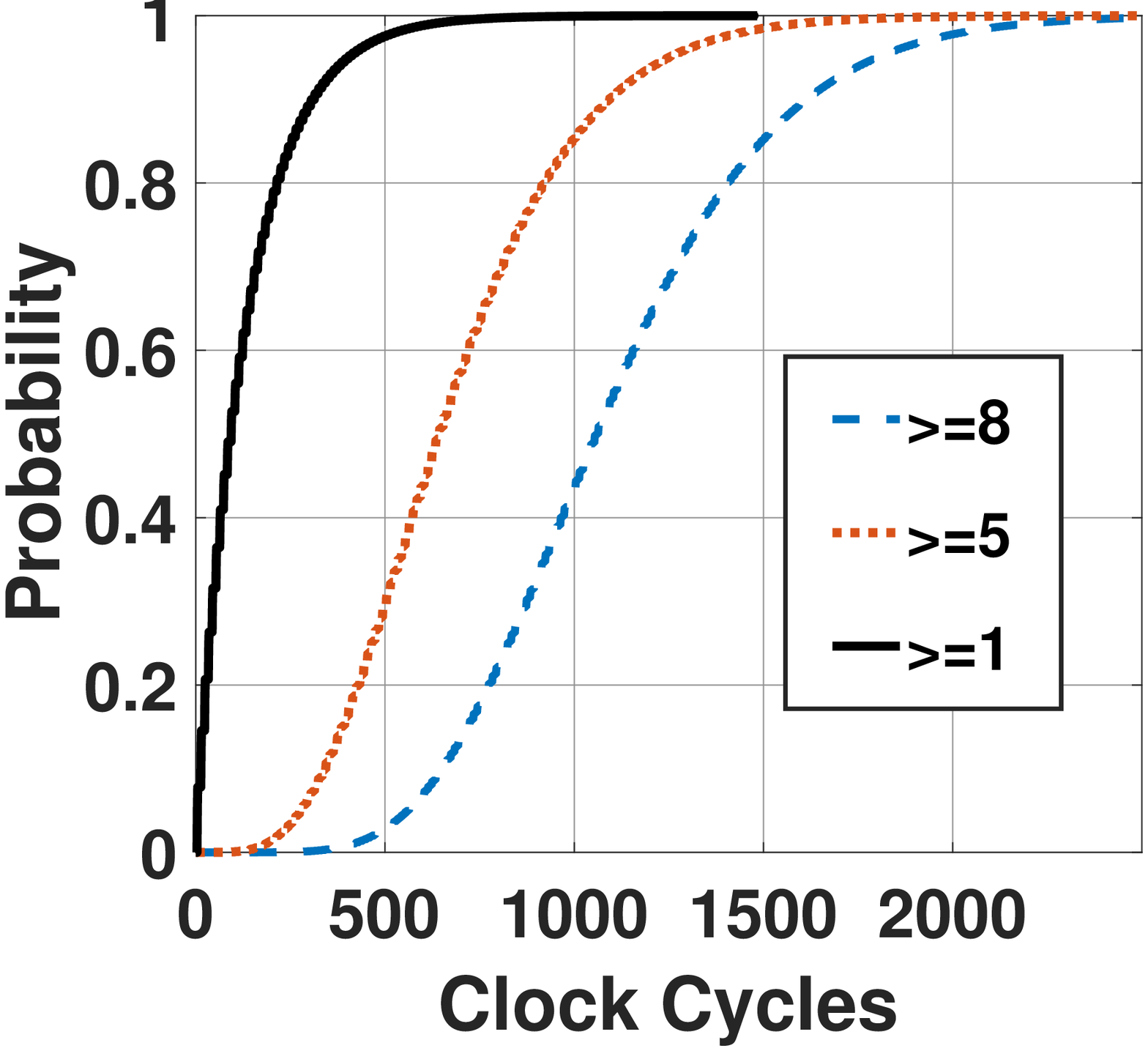}
    }
    \caption{CDF for \noiserun}
\end{subfigure}
    \caption{CDF for \optrun and \noiserun in Burst of Three Input Configuration}
    \label{fig:BurstNoiseRes}
\end{figure}

This input configuration results in only 3 flits every 10 cycles for each arbiter to service in a router, as opposed to 5 for the every-other-cycle injection. However, the ability of each arbiter to receive flits during consecutive cycles has a major impact on the PSN behavior of the NoC. The likelihood of having both \optrun and \noiserun increases significantly. Figure~\ref{fig:BurstNoiseRes} shows the plots of the \emph{cumulative distribution function} (CDF) for \optrun being greater than or equal to 1, 5, 10, and 20, and the CDF for \noiserun being greater than or equal to 1, 5, and 8. Comparing these \optrun and \noiserun probabilities with that of the every other cycle injection shows that PSN is much more likely with bursty style injection, within the same number of clock cycles, despite fewer packets entering the NoC every 10 cycles.

\subsection{Flit Generation Pattern to Minimize PSN}
We parameterized the bursty mode model so that it could accept a burst lasting any number of cycles, followed by any number of idle cycles. Under our memory constraints, we were able to model check bursts lasting 1, 2, or 3 cycles long, with the requirement that the number of idle cycles must be at least twice as long as the number of cycles in the burst. When testing various burst configurations, we made a critical observation that the input configuration of 1 flit every 3 cycles incurred \emph{zero} probability of an \optrun event, and, by extension, zero probability of an \noiserun event. This is of particular note, because 1 every 3 flit injection results in more flits being injected over time than 3 every 10 injection, but with no occurrence of high-PSN events. Additionally, 1 every 3 flit injection allows considerable number of cycles to be verified compared to every other cycle flit generation and yet it still incurred zero probability of PSN.

\subsection{Results Summary and Discussion}
The experiments on every other cycle injection suggest that applying PMC on the Boolean queue abstract model is the only viable option to verify PSN related probabilistic properties at longer cycles with reasonable accuracy, given the rarity of the properties. Verification results indicate extremely low probabilities in observing a \optrun event within 30 clock cycles. Since \noiserun accumulates cycles with high-to-low or low-to-high activities, rare occurrences of \optrun consequently lead to extremely low probabilities in \noiserun verification. 
On the other hand, PMC on the 3 every 10 burst flit injection scales PMC to allow much longer cycles and results in considerably higher PSN due to significantly increased \optrun and \noiserun probabilities. Under our memory constraints, 1 of every 3 flit injection is the most effective packet injection pattern that minimizes PSN probability to zero.

The drastically different PSN behaviors can be explained by analyzing local buffer and arbiter activities of each router.
Since the entire NoC only includes four corner routers, each flit has a relatively short distance to its destination router, which reduces the number of buffers it has to visit. Therefore, it is unlikely that all three buffers can contain a flit at the same time, a condition necessary for a \optrun cycle to occur. In addition, with X-Y routing, flits tend to exit the NoC quickly without filling up buffers in a way that is conducive to \optrun. For the burst mode flit injection, because all three buffers in a router simultaneously start three cycles of burst of flits, it is more likely to incur high router activities and switching between high and low router activities. The burst mode is more prone to cycles of high activity when the NoC has consecutive cycles of flit injection, because it does not have cycles to clear out the buffers before another injection. This causes more traffic in the buffers during the burst, leading to more cycles of high router activity overall, despite the seven cycles of idle behavior. An optimal compromise is to have 1 flit every 3 cycles where the occurence of the high PSN events we are tracking reduces to none.

Our findings indicate that spreading flits over a small number of cycles, rather than releasing them in consecutive ones, can drastically reduce PSN. In this work, we made assumption to allow each router to generate flits. In real-world NoC design, network flits are generated and scheduled by external components. Traditional techniques (e.g., IcoNoClast\cite{basu:iconoclast:tvlsi17}) enhance the router microarchitecture to delay the traversal of flits within a network, thus effectively curbing the maximum noise of the communication fabric. However, such schemes incur additional design complexity and hardware overheads that could be prohibitive for low-power edge applications. Based on our PMC approach, we set a more cost-effective approach to tackling the communication noise in the low-power domains. Our findings suggest that while scheduling network flits, the scheduler should try to insert empty cycles to separate flits in succession in order to minimize PSN.

 
\section{Conclusion}

This paper describes our experience in formally modeling a 2$\times$2 NoC system and applying probabilistic verification to quantitatively verify the frequency of PSN. 
Probabilistic model checking using \mcsta was used to evaluate the properties. The concrete model incurs severe state explosion that prevents property checking. Several abstraction techniques, including a novel probabilistic choice abstraction, are applied to alleviate the rapid state-space explosion and allow for successful verification. Results indicate that bursty flit injection in consecutive cycles yields high likelihood of PSN-causing behavior, while spreading flits over a small number of cycles achieves PSN reduction. 
This shows that flit injection patterns affect PSN drastically, and that by using certain flit injection patterns, such as 1 flit every 3 cycles, more flits can be injected while also minimizing PSN. For further work, we plan to investigate methods to make the use of BDDs more successful in analyzing this NoC. Additionally, we plan to extend the PSN analysis to larger NoCs, and with different flit generation rates. In order to scale model checking, we plan to pursue the use of assume-guarantee reasoning to combat state explosion. Additionally, we plan to work on a formal proof for the probabilistic choice abstraction. 

\subsubsection*{Acknowledgments. }
Arnd Hartmanns was supported by NWO VENI grant 639.021.754. Riley
Roberts, Benjamin Lewis, Koushik Chakraborty, and Sanghamitra Roy were
supported in part by National Science Foundation (NSF) grants {\small
  CAREER-1253024}, {\small CNS-1421022}, and {\small CNS-1421068}. Any
opinions, findings, and conclusions or recommendations expressed in
this material are those of the authors and do not necessarily reflect
the views of the~NSF. Riley Roberts and Benjamin Lewis were supported
in part by gift donations from Adobe Systems Incorporated.

\bibliographystyle{IEEEtran}{
\bibliography{main}
}

\end{document}